\documentclass[aps,prl,reprint,superscriptaddress,floatfix]{revtex4-2}
\usepackage{amssymb,graphicx}
\usepackage[dvipsnames,usenames]{xcolor}
\usepackage[colorlinks,citecolor=blue]{hyperref}

\begin{document}
\title{Production of $p$-nuclei from $r$-process seeds: the $\nu r$-process}
\author{Zewei Xiong}
\email[Email: ]{z.xiong@gsi.de}
\affiliation{GSI Helmholtzzentrum {f\"ur} Schwerionenforschung, Planckstra{\ss}e 1, D-64291 Darmstadt, Germany}

\author{Gabriel Mart\'{i}nez-Pinedo}
\affiliation{GSI Helmholtzzentrum {f\"ur} Schwerionenforschung, Planckstra{\ss}e 1, D-64291 Darmstadt, Germany}
\affiliation{Institut {f\"ur} Kernphysik (Theoriezentrum), Fachbereich Physik, Technische Universit{\"a}t Darmstadt, Schlossgartenstra{\ss}e 2, D-64289 Darmstadt, Germany}
\affiliation{Helmholtz Forschungsakademie Hessen {f\"ur} FAIR, GSI Helmholtzzentrum {f\"ur} Schwerionenforschung, Planckstra{\ss}e 1, D-64291 Darmstadt, Germany}

\author{Oliver Just}
\affiliation{GSI Helmholtzzentrum {f\"ur} Schwerionenforschung, Planckstra{\ss}e 1, D-64291 Darmstadt, Germany}
\affiliation{Astrophysical Big Bang Laboratory, RIKEN Cluster for Pioneering Research, 2-1 Hirosawa, Wako, Saitama 351-0198, Japan}

\author{Andre Sieverding}
\affiliation{Max Planck Institute for Astrophysics, Karl-Schwarzschild-Stra{\ss}e 1, D-85748 Garching, Germany}

\date{\today}

\begin{abstract}
We present a \emph{new} nucleosynthesis process that may take place on neutron-rich ejecta experiencing an intensive neutrino flux.
The nucleosynthesis proceeds similarly to the standard $r$-process, a sequence of neutron-captures and beta-decays, however with charged-current neutrino absorption reactions on nuclei operating much faster than beta-decays.
Once neutron capture reactions freeze-out the produced $r$-process neutron-rich nuclei undergo a fast conversion of neutrons into protons and are pushed even beyond the $\beta$-stability line producing the neutron-deficient $p$-nuclei.
This scenario, which we denote as the $\nu r$-process, provides an alternative channel for the production of $p$-nuclei and the short-lived nucleus $^{92}$Nb.
We discuss the necessary conditions posed on the astrophysical site for the $\nu r$-process to be realized in nature.
While these conditions are not fulfilled by current neutrino-hydrodynamic models of $r$-process sites, future models, including more complex physics and a larger variety of outflow conditions, may achieve the necessary conditions in some regions of the ejecta.
\end{abstract}

\maketitle
\graphicspath{{./}{./figures/}}

\paragraph{Introduction}

A variety of processes have been suggested as the origin of stable nuclei heavier than iron and located at the neutron-deficient side of the $\beta$-stability line, the so-called $p$-nuclei.
Those include the $\gamma$-process ($p$-process)~\cite{arnould2003p,pignatari2016production}, $\nu p$-process~\cite{frohlich2006neutrino,Pruet.Hoffman.ea:2006,wanajo2006rp}, and $rp$-process~\cite{Schatz.Aprahamian.ea:1998}.

In the $\gamma$-process, seed nuclei present from the initial composition of the star, undergo $(\gamma,n)$ reactions followed by $(\gamma,p)$ or $(\gamma,\alpha)$ as the temperature rises to 3--5~GK in core-collapse and Type Ia supernovae~\cite{hayakawa2004evidence,kusakabe2011production,Travaglio.Roepke.ea:2011,pignatari2016production}.
While its yields are consistent for more than half of the $p$-nuclei, some specific regions, such as $^{92,94}$Mo and $^{96,98}$Ru, are underproduced.
The $\gamma$-process is not a primary process and depends on the preexisting $s$-process seeds.

On the other hand, $p$-nuclei can also be produced in the $\nu p$-process through proton capture aided by neutrinos in neutrino-driven winds from core-collapse supernovae (CCSNe)~\cite{frohlich2006neutrino,wanajo2011uncertainties}.
Long $\beta^+$ decay times of waiting-point nuclei such as $^{64}$Ge can be circumvented by $(n,p)$ reactions with neutrons produced by absorption of electron antineutrinos on protons.
However, current three-dimensional supernova models suggest that a neutrino-driven wind may not develop except for low mass progenitors~\cite{Bollig.Yadav.ea:2021}.
Furthermore, neutrino-wind simulations based on up-to-date set of neutrino opacities produce ejecta not proton-rich enough for a strong $\nu p$-process~\cite{Fischer.Guo.ea:2020}.

Light $p$-nuclei, like $^{92}$Mo, may also be produced in the inner ejecta of CCSN by explosive nucleosynthesis~\cite{hoffman1996production,wanajo2018nucleosynthesis,sieverding2020nucleosynthesis}, but no substantial production occurs of heavier ones.
Light $p$-nuclei can also be produced by the $rp$-process in accreting neutron stars~\cite{Schatz.Aprahamian.ea:1998}, but it is an open question whether and how material is ejected from the neutron star and contributes to galactic chemical evolution.

In addition to $p$-nuclei, another element of yet unknown origin is the by now extinct nucleus $^{92}$Nb that existed in the early solar system (ESS) \cite{dauphas2003short,dauphas2011perspective,hayakawa2013supernova,hibiya2023evidence}.
It cannot be produced by the $\nu p$- or $r p$-processes as it is shielded by $^{92}$Mo~\cite{rauscher2013constraining}.
The production of $^{92}$Nb through the $\gamma$-process is viable but the yield falls short of explaining the amount measured in the ESS, probably related to the underproduction of $^{92,94}$Mo and $^{96,98}$Ru~\cite{lugaro2016origin}.
A feasible way of production could be through charge-current (CC) and neutral-current (NC) weak interactions on the preexisting nuclei $^{92}$Zr and $^{93}$Nb in the $\nu$-process \cite{hayakawa2013supernova,sieverding2018nu,sieverding2019nu}.

Previous studies of the $r$-process both in the context of CCSNe and binary neutron star mergers (BNM) have shown that neutrinos play a fundamental role.
At high temperatures, when the composition consists of neutrons and protons, electron (anti)neutrino absorption and the inverse reactions determine the neutron-richness of the ejected material~\cite{Arcones.Thielemann:2013,goriely2015impact,martin2018role,cowan2021origin,Kullmann.Goriely.ea:2022}.
This aspect is fundamental to produce ejecta with a broad distribution of neutron-richness and to account for the observation of Sr in the AT2017gfo kilonova~\cite{Watson.Hansen.ea:2019}.
Large neutrino fluxes are in general detrimental for the $r$-process as they drive the composition to proton-to-nucleon ratios of $Y_e \approx 0.5$ due to the $\alpha$-effect~\cite{bradley1998neutrino,langanke2003nuclear}.
Once nuclei form, substantial neutrino fluxes hinder the operation of the $r$-process by converting neutrons into protons and reducing the amount of neutrons available for captures.
Furthermore, large rates of electron neutrino absorption on nuclei hinder the formation of $r$-process peaks associated to magic numbers~\cite{fuller1995neutrino,Mclaughlin.Fuller:1996,qian1997neutrino}.

This regime of large neutrino fluxes, when electron-neutrino absorption rates are faster than beta-decays for neutron-rich nuclei, is precisely the regime we consider in this letter.
We will show that it leads to a \emph{new} nucleosynthesis process that produces $p$-nuclei and $^{92}$Nb operating on seeds produced by the $r$-process under strong irradiation by neutrinos.
We denote this process as $\nu r$-process.
Unlike $\gamma$-, $\nu p$-, or $r p$-processes that occur in proton-rich conditions, the $\nu r$-process operates in neutron-rich conditions.
The large neutrino fluxes restrict the conditions at which it operates to $Y_e \approx 0.4$--0.5.
It requires several stages for the production of $p$-nuclei.
The nucleosynthesis starts at high temperatures with a composition of neutrons and protons.
As the temperature decreases, nuclei are formed by charged-particle reactions that freeze out at temperatures $T\sim 3$~GK.
This phase results in an $\alpha$-rich composition characteristic of the operation of the $\alpha$-process~\cite{woosley1992alpha,Witti.Janka.Takahashi:1994}.
The produced nuclei act as seeds for further neutron captures along a path determined by $(n,\gamma)\rightleftarrows(\gamma,n)$ equilibrium. 
In the absence of neutrino irradiation, the equilibrium is broken by beta-decays that determine the speed at which heavy elements are build and the rate at which neutrons are exhausted.
When this happens, neutron-captures freeze-out and nuclei undergo beta-decay and migrate towards the valley of stability.

This picture, however, is drastically changed if the ejecta are irradiated by a large neutrino flux.
Large neutrino-absorption rates dominate the flow to higher charge numbers instead of beta-decays, leading to faster depletion of neutrons.
This speeds up the production of heavy elements and leads to an earlier freeze-out of neutron captures at higher temperatures compared with the case without neutrino fluxes.
Unlike the $r$-process, in which after the freeze-out material moves to the stability by beta-decays stopping at the first stable nucleus,  in the $\nu r$-process CC neutrino-nucleus reactions drive the matter to and beyond the stability line.  
Furthermore, the energy of neutrinos is larger than the neutron, proton and alpha separation energies, and hence CC and NC neutrino-nucleus reactions produce free neutrons, protons and $\alpha$-particles that are captured at the still relatively high temperatures.
This leads to a broader abundance distribution that reaches from neutron-deficient to neutron-rich nuclei on both sides of the stability line and the production of significant amounts of $p$-nuclei.

\paragraph{Parametric trajectory}
We use the nuclear reaction network (employed previously in, e.g., Refs.~\cite{collins2023radiative,just2023end}) with the reaction rates based on the finite-range droplet mass model \cite{mendoza2015nuclear} and a consistent description of neutrino reactions with nucleons ($\nu$-N) and nuclei ($\nu$-A) that considers light particle spallation induced by both NC and CC $\nu$-A reactions~\cite{sieverding2018nu}.
We neglect finite temperature corrections to the neutrino-nucleus cross sections.

We assume adiabatic expansions starting with an initial temperature of $T_0=10$~GK and density of $\rho_0 = 4.6\times 10^6$~g~cm$^{-3}$ that corresponds to an initial entropy per nucleon, $s=84$~$k_B$.
We assume homologous expansion for the density evolution $\rho(t)=\rho_0 (1+t/\delta_\rho)^{-3}$.
The temperature is evolved accounting for the energy generation by nuclear reactions~\cite{Reichert.Winteler.ea:2023}.
$\delta_\rho$ is related to the expansion timescale $\tau_{\text{exp}} = -[d\ln\rho(t)/dt]^{-1} = (t+\delta_\rho)/3$.
We fix the initial electron fraction, $Y_{e,0}$, that is subsequently evolved under the influence of (anti)neutrino absorption and their inverse reactions.
Neutrino number densities are parameterized as $n_{\nu_e}(t) = n_{\nu_e,0} (1+t/\delta_\nu)^{-3}$ and $n_{\bar\nu_e}(t)=R_n n_{\nu_e}(t)$, where $\delta_\nu\geq \delta_\rho$ (because the neutrino flux may decrease more slowly than the baryon density at small radii, e.g., due to polar focusing of radiation in an accretion disk \cite{just2015comprehensive}), and $R_n$ is a constant ratio relating the $\bar{\nu}_e$ and $\nu_e$ densities.
The power index $-3$ accounts for both inverse-square radial dependence of the neutrino flux at $t\gtrsim \delta_\nu$ and the decay of the neutrino luminosity assumed here $\propto t^{-1}$.
Neutrino spectra are approximated by Fermi-Dirac distributions with constant effective temperatures $T_{\nu_e}$ and $T_{\bar\nu_e}$.

\begin{figure}[hbt]
  \centering
  \includegraphics[width=\columnwidth]{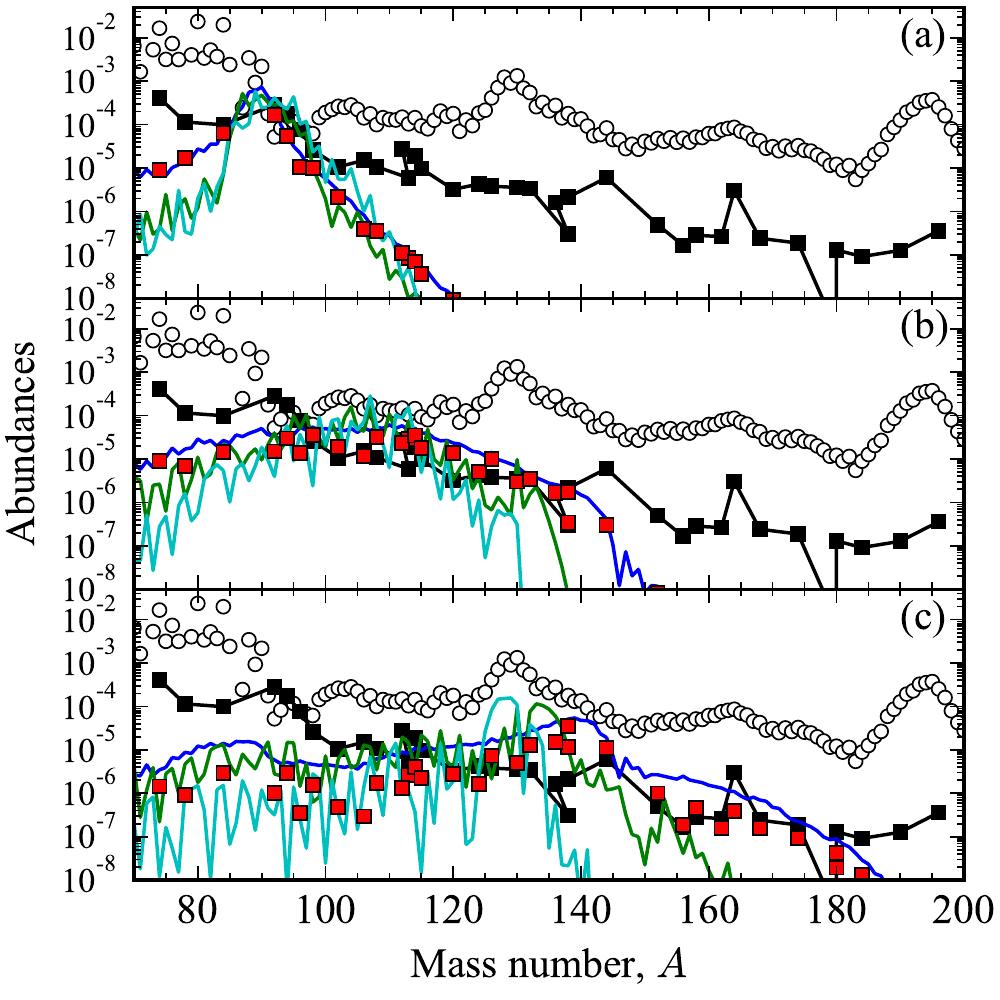}
  \caption{\label{fig:abundance}Comparison of the isobaric yields at 1~Gyr (blue curves) and the abundance of $p$-nuclei on those yields (red squares) with the solar abundances of $p$-nuclei (black squares)~\cite{arnould2003p} and $r$-nuclei (black circles)~\cite{Goriely:1999} in cases I (a), II (b), and III (c).
  Green and cyan curves show the yields at $n_s\sim1$ with and without $\nu$-A reactions, respectively.}
\end{figure}

Figure~\ref{fig:abundance} compares the abundances from three trajectories I--III with parameters $[\delta_\rho\,~(\text{ms}), n_{\nu_e,0}\, (10^{32}\ \text{cm}^{-3})]$ of $[4, 2.5]$, $[1, 10]$, and $[0.5, 20]$, respectively.
The other parameters have the values: $Y_{e,0}=0.4$, $R_n=1.2$, $\delta_\nu=4\delta_\rho$, $T_{\nu_e}=5$~MeV, and $T_{\bar\nu_e}=1.25 T_{\nu_e}$.
We find that $Y_e$ reaches $\approx 0.467$ at $T=5$~GK in all three cases.
At this temperature, the rapid expansion of moderately neutron-rich matter gives high neutron-to-seed ratios $n_s$ (13.8, 59.2, and 177, respectively), and allows an $r$-process to occur.

\paragraph{Abundances}
Without $\nu$-A reactions, case I corresponds to the $\alpha$-rich freeze-out with moderate neutron-rich conditions, such as considered in Refs.~\cite{hoffman1996production,wanajo2006rp,Bliss.Arcones.Qian:2018}, that is known to produce $^{92}$Mo but not $^{94}$Mo.
Once the $\nu$-A reactions are included, the ratio $^{92}$Mo/$^{94}$Mo becomes consistent with solar proportions and the yield of the radioactive $^{92}$Nb is also enhanced.
Despite their much larger $n_s$ at 5~GK that would usually allow for a much stronger $r$-process, the abundance peaks for cases II and III only moderately shift to higher mass numbers when compared to case I.
The larger neutrino densities lead to a conversion of neutrons into protons while the $r$-process operates and reduce the amount of neutrons for captures on heavy nuclei.
When $n_s\sim 1$ is reached, the initial supply of neutrons is almost depleted.
Figure~\ref{fig:abundance} compares the abundances at this time with and without $\nu$-A interactions (cyan vs. green lines), illustrating that $\nu$-A interactions speed up the flow to heavier isotopes.
The difference between the abundances at $n_s\sim 1$ and the final abundances shows the impact of the additional neutrons produced by the neutrino spallation reactions.

All those three cases show comparable amounts of $p$-nuclei produced relative to the final total yield at mass numbers close to the respective abundance peak, which indicates a successful conversion of those $r$-process seeds.
The abundance pattern in case I shows a peak around $^{92,94}$Mo and $^{96,98}$Ru.
Larger $n_s$ in cases II and III lead to the production of $p$-nuclei up to $A\sim 145$ and 180, respectively.
Heavier $p$-nuclei up to $^{196}$Hg can be produced with higher neutrino number densities.

\begin{figure}[hbt!]
	\centering
    \includegraphics[width=\columnwidth]{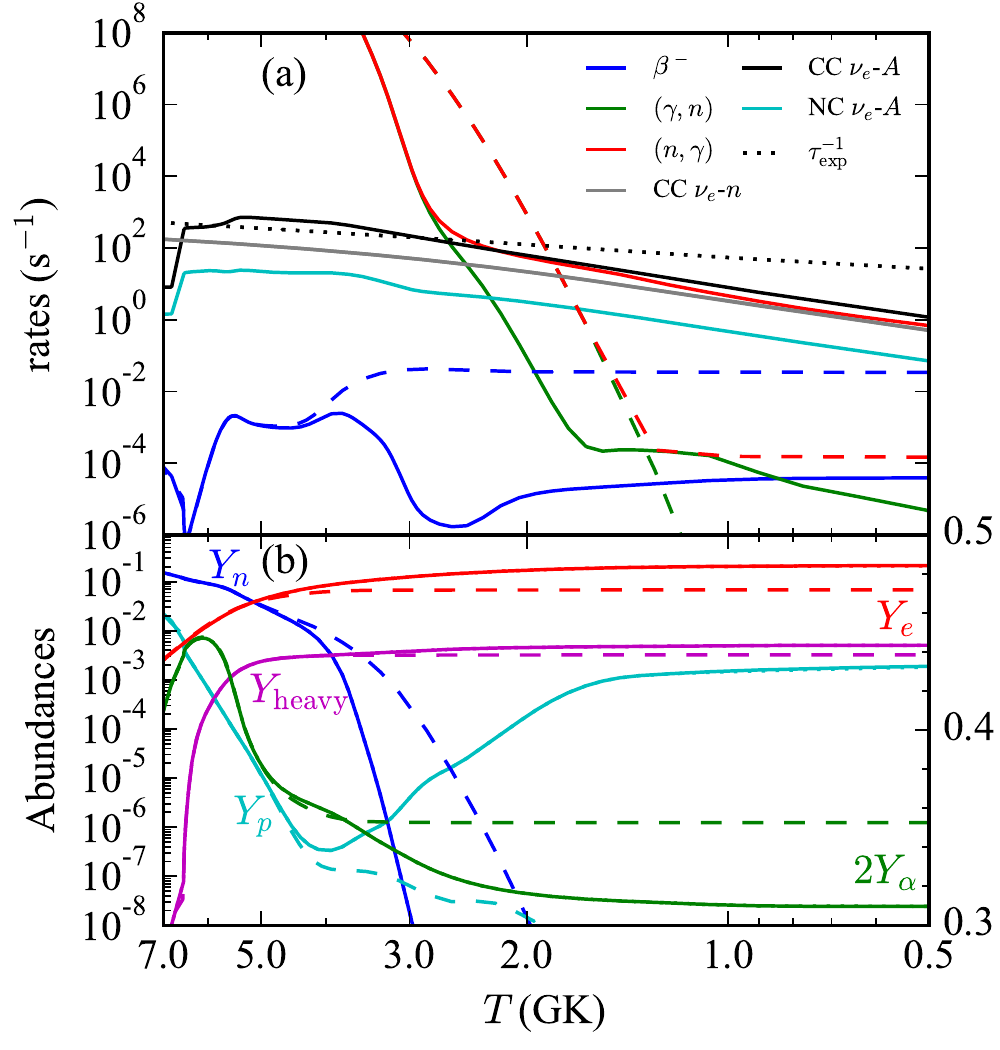}
    \caption{\label{fig:rate}Average rates (a) and abundances (b) as functions of temperature including (solid) and neglecting (dashed) $\nu$-A reactions in case I.
   $Y_\alpha$ and $Y_e$ are shown with the scale labeled on the right side.}
\end{figure}

\paragraph{Reaction dynamics}
We compare the averaged rates for $\beta^-$ decay, $(n,\gamma)$, $(\gamma,n)$,
CC and NC $\nu_e$-A reactions for case I in Fig.~\ref{fig:rate}(a).
They are computed as $\lambda_{I} = \sum_i \lambda_{I, i} Y_i/ Y_{\text{heavy}}$ where $I$ stands for a particular reaction, $i$ sums over all heavy nuclei ($A>4$), and $Y_{\text{heavy}}=\sum_i Y_i$ is the total abundance of heavy nuclei.
The $\nu_e$-$n$ rate $\lambda_{\nu_e n} = n_{\nu_e} \sigma_{\nu_e n}\, c$ is approximately 1~ms$^{-1}$ for $n_{\nu_e}=10^{33}$~cm$^{-3}$ and $T_{\nu_e}=5$~MeV.
The CC rates $\lambda_{\nu_e A,i}^{\text{CC}}= n_{\nu_e} \sigma_{\nu_e A,i}^{\text{CC}}\, c$ for nuclei with $A=100$--200 near the stability line are $\sim 1$--10 times larger than $\lambda_{\nu_e n}$.

When the temperature is above $\sim 4.5$~GK, there is essentially no difference between the cases with and without $\nu$-A reactions.
As the temperature becomes lower, and the composition shifts to neutron-rich nuclei, $\lambda_{\nu_e A}^{\text{CC}}$ becomes comparable with the expansion rate leading to a faster depletion of neutrons because $\nu_e$ absorption on nuclei speeds up the production of heavy elements.
The balance between $(n,\gamma)$ and $(\gamma,n)$ is broken at $\sim 3$~GK.
The rate of $(\gamma,n)$ decreases drastically, but the rate of $(n,\gamma)$ changes gradually as neutrons are continuously supplied from the neutrino spallation by both CC and NC $\nu_e$-A reactions.
The $(n,\gamma)$ rate follows $\lambda_{\nu_e A}^{\text{CC}}$ and the nucleosynthesis flow is  characterized by an equilibrium between $(n,\gamma)$ and $\nu_e$-A reactions.
This produces a rather broad abundance distribution that reaches from the neutron-deficient to the neutron-rich side of beta-stability.
Since more neutron-rich isotopes are reached, the average $\beta^-$ rate increases slightly below $T\sim 3$~GK.
The average number of emitted neutrons from the CC $\nu$-A reaction is approximately given by the ratio $\lambda_{(n,\gamma)}/\lambda_{\nu_e A}^{\text{CC}}$.
(The average rate of NC reactions is about one order of magnitude lower than $\lambda_{\nu_e A}^{\text{CC}}$.)
We note that we do not consider heavy lepton neutrinos.
They have higher mean energy and possibly larger flux and hence may amplify the total rate to values comparable to $\lambda_{\nu_e A}^{\text{CC}}$.

Neutrinos continue to interact with heavy nuclei and increase $Y_e$ as shown in Fig.~\ref{fig:rate}(b).
Neutrino spallation on $\alpha$-particles produces $^3$H and $^3$He that are converted into $^{12}$C by additional $\alpha$-captures.
This is achieved by two-body reactions instead of the triple-alpha reaction, and hence they operate till much lower temperatures.
After the production of $^{12}$C, a sequence of $\alpha$-captures continues, similarly to the $\alpha p$- and $rp$-process on accreting neutron stars~\cite{schatz2006x}.
As a result, $Y_\alpha$ decreases and $Y_{\rm heavy}$ increases.
Protons produced from neutrino-spallation are barely used below $\sim4$~GK.
This together with their production by $(n,p)$ reactions increases $Y_p$ to values $\sim 10^{-3}$.
See more details in the Supplemental Material \cite{SM}.

\begin{figure}[hbt!]
  \centering
  \includegraphics[width=\columnwidth]{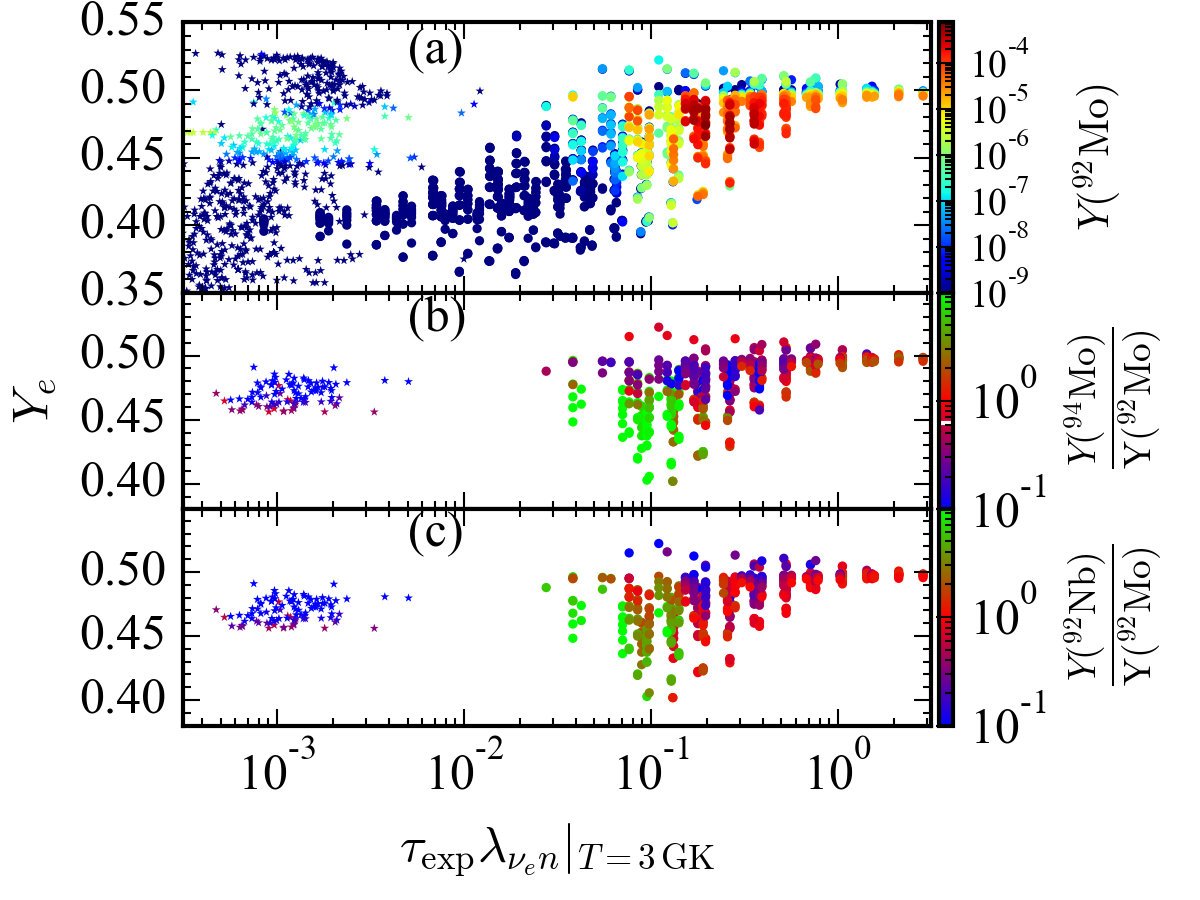}
  \caption{\label{fig:exposure}Scatter plots for the abundance of $^{92}$Mo (a), the ratio $^{94}\text{Mo}/^{92}\text{Mo}$ (b), and the ratio $^{92}\text{Nb}/^{92}\text{Mo}$ (c) with respect to $Y_e$ and neutrino exposure from our parametric survey (circles) and the simulation model sym-n1-a6 (stars) from Ref.~\cite{just2023end}.
  The white line in the color bar in (b) shows the $^{94}\text{Mo}/^{92}\text{Mo}$ ratio in solar abundance.
  (b) and (c) show the abundance ratio only for $Y(^{92}\text{Mo})>10^{-7}$.}
\end{figure}

\paragraph{Survey of conditions}
In addition to the previous three cases, we surveyed a larger parameter space with the following variations: $\delta_\rho=\{0.5,1,2,4,8\}$~ms, $n_{\nu_e,0}=\{0.25,0.5,1,2\}\times 10^{33}$~cm$^{-3}$, $R_n=\{0.5$,0.8,1,1.2,1.5,2,$4\}$ and $\delta_\nu=\{1,1.5,2.5,4,6\}\times \delta_\rho$, while keeping others the same.
We classify the trajectories by the value of $Y_e$ reached at 5~GK and the neutrino exposure.
As a measure of the latter, we use the product of the expansion timescale and the rate for $\nu_e$ absorption on neutrons, both evaluated at 3~GK.
The three panels of Fig.~\ref{fig:exposure} show the abundance of $^{92}$Mo, the abundance ratio $^{94}$Mo/$^{92}$Mo, and the ratio $^{92}$Nb/$^{92}$Mo at 1~Myr, respectively.
For low values of the neutrino exposure all calculations are concentrated around $Y_{e}\approx 0.4$ because of our choice of $Y_{e,0}$.
High neutrino exposure makes all calculations converge to $Y_e\approx 0.5$.
We observe large variations of the abundance yields in the intermediate region.
Significant amounts of $^{92}$Mo ($\gtrsim 10^{-4}$, corresponding to more than 2.5\% of the heavy-element mass when the alpha mass fraction is $\approx 0.6$) are produced when $\tau_{\text{exp}}\lambda_{\nu_e n}\gtrsim 0.1$.
The ratios $^{94}$Mo/$^{92}$Mo and $^{92}$Nb/$^{92}$Mo are larger for the neutron-rich cases with small neutrino exposure, and are below unity when $Y_e\approx 0.5$.

Figure~\ref{fig:exposure} also shows the nucleosynthesis results for an exemplary neutrino-hydrodynamics simulation of a BNM, namely model sym-n1-a6 of Ref.~\cite{just2023end}.
In this simulation, neutrino fluxes become negligible by the time nuclei form.
Consequently, $p$-nuclei around $A\sim 92$ are produced by the $\alpha$-process operating in a narrow window of $Y_e \sim 0.45$--0.48
while $p$-nuclei heavier than $^{92}$Mo are underproduced.
However, the conditions in current models are not very far off, and the desired neutrino exposure can be achieved for the same trajectories at earlier times when $T=10$~GK as shown in the Supplemental Material \cite{SM}.
The comparison suggests that although purely neutrino-driven winds are not well suited for the $\nu r$-process, outflows not driven by thermal expansion may be more promising, because their temperature is lower close to the central object with high neutrino fluxes.
We speculate that the $\nu r$-process may operate in magnetically-driven ejecta subject to strong neutrino fluxes, which are likely found in polar regions of magnetorotational supernovae~\cite{Obergaulinger.Reichert:2023} and collapsar engines~\cite{Siegel.Barnes.Metzger:2019}, or around magnetized remnants of BNMs~\cite{Miller.Ryan.ea:2019,Shibata.Fujibayashi.Sekiguchi:2021,moesta2020magnetar}.

\paragraph{Observables}
Assuming that the $\nu r$-process produces $p$-nuclei in an astrophysical site where also the $r$-process operates and that the yields of both follow solar proportions, we can obtain constraints on the relative contributions of the $\nu r$-process and the $r$-process to the ejecta.
If the whole solar inventory of $p$- and $r$-nuclei is produced on the same site, we expect a ratio of $\sim 90/1$ between the $r$-process and $\nu r$-process material.
We notice that the observation of Sr in the kilonova transient AT2017gfo and the low lanthanide mass fraction inferred from multiband light-curve analyses requires the production of all $r$-process nuclei~\cite{Wu.Barnes.ea:2019}, assuming solar proportions. 

\begin{figure}[hbt]
	\centering
    \includegraphics[width=\columnwidth]{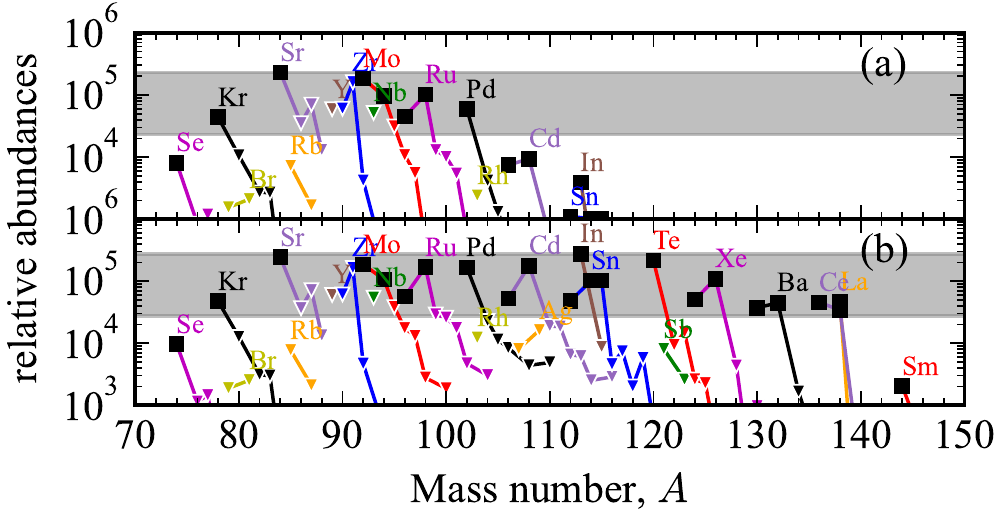}
    \caption{\label{fig:relabundance}Relative ratios of $p$-nuclei (black squares) and other isotopes (colored triangles) over the solar abundance \cite{asplund2009chemical} for case I (a) and case I plus 20\% of case II (b).
    Nuclei in the gray bands have a relative abundance within an order of magnitude of the maximum.}
\end{figure}

We show in Fig.~\ref{fig:relabundance}(a) the production factors of $p$-nuclei, i.e. the abundances normalized to the solar value $Y_i/Y_{i,\odot}$ for case I.
The gray band, which covers one order of magnitude right below the largest abundance, illustrates the isotopes that are expected to be co-produced.
The main productions of $p$-nuclei in cases I and II are from $^{78}$Kr to $^{102}$Pd and from $^{98}$Ru to $^{138}$La, respectively.
Given that case II requires very high neutrino fluxes, we expect that it is more rare than case I.
Therefore, we combine 20\% from case II with 80\% from case I in Fig.~\ref{fig:relabundance}(b), which yields a pattern that is in good agreement with the solar abundances of the $p$-nuclei from $^{78}$Kr to $^{138}$La.
The $\nu r$-process does not only produce nuclei commonly associated to the $\gamma$-process, but also $^{138}$La and $^{180}$Ta that are often associated with the $\nu$-process~\cite{sieverding2018nu}.

Associating the origin of the $p$-nuclei with an $r$-process site, such as BNMs, may have implications for the correlated excesses between $p$- and $r$-nuclei of Mo in primitive meteorites~\cite{stephan2019molybdenum}.
We can further estimate the time since the last $\nu r$-process addition to the solar system by considering the short-lived radioactive isotope $^{92}$Nb.
The $\nu r$-process results in an abundance ratio between $^{92}$Nb and $^{92}$Mo that is close to unity, as shown in Fig.~\ref{fig:exposure}(c), which is $\sim 10^3$ larger than the ratio $\sim \mathcal{O}(10^{-3})$ found from the $\nu$-process in CCSNe \cite{hayakawa2013supernova,sieverding2018nu}.
Assuming a production ratio close to unity in a simple model of uniform production over the age of the universe~\citep{wasserburg2006short,huss2009stellar} and supposing that both, $^{92}$Mo and $^{92}$Nb are predominantly made from the $\nu r$-process, the ratio in the interstellar medium is $\sim 5 \times 10^{-3}$ after 10~Gyr of evolution.
This rough estimate suggests a decay time of $\sim 250$~Myr since the last event to match the ESS ratio, which is consistent with the expected time of $\sim 100$~Myr since the last $r$-process event \cite{lugaro2014chronometer,lugaro2018radioactive}.

Since the $\nu r$-process depends on the neutrino flux, it could be significantly affected by collective neutrino flavor phenomena \cite{duan2011influence,xiong2020potential,george2020fast,li2021neutrino,just2022fast,fernandez2022fast}.
In addition, the process depends on the competition between neutrino absorption and neutron captures near the stability line, which calls for further improved measurements of the reaction rates in the laboratory.
In the present work, we have considered moderately neutron-rich ejecta and shown that large neutrino fluxes can drive the composition to the neutron-deficient site of the stability valley producing $p$-nuclei, not only for Mo and Ru but also heavier ones.
One can wonder if a reverse process may operate in proton-rich ejecta driving the composition from the neutron-deficient side to the neutron-rich side by $\bar{\nu}_e$-A reactions and producing $r$-nuclei.
We expect that this will not be the case as neutron-deficient nuclei still have a neutron excess and hence similar or even larger cross sections for $\nu_e$-A absorption than $\bar{\nu}_e$-A absorption.
It remains to be explored if variations of the neutrino flux can produce abundance patterns that resemble those of the $s$-process or $i$-process.

\vspace{0.1in}
\textit{Note added.}—--A recent study \cite{prasanna2024favorable} has reported that suitable conditions for the $\nu r$-process may exist in winds from rotating highly magnetized proto-neutron stars.
\vspace{0.1in}

\begin{acknowledgments}
We thank Andreas Bauswein, Karlheinz Langanke, Yong-Zhong Qian,  Ninoy Rahman, and Friedrich-Karl Thielemann for fruitful discussions.
GMP and ZX acknowledge support by the European Research Council (ERC) under the European Union's Horizon 2020 research and innovation programme (ERC Advanced Grant KILONOVA No.~885281).
OJ acknowledges support by the ERC under grant agreement No. 759253.
OJ, GMP, and ZX acknowledge support by the Deutsche Forschungsgemeinschaft (DFG, German Research Foundation) - Project-ID 279384907 - SFB 1245, and MA 4248/3-1.
GMP and OJ acknowledge support by the State of Hesse within the Cluster Project ELEMENTS.
GMP thanks the hospitality of the Instituto de Física Teórica UAM-CSIC, supported by the Severo Ochoa Excellence Program No CEX2020-001007-S funded by MCIN/AEI, where part of this work was done.
AS acknowledges funding from the European Union’s Framework Programme for Research and Innovation Horizon Europe under Marie Sklodowska-Curie grant agreement No. 101065891.
\end{acknowledgments}

\bibliography{main.bbl}

\end{document}


\title{Supplemental Material for ``Production of $p$-nuclei from $r$-process seeds: the $\nu r$-process''}
\author{Zewei Xiong}
\email[Email: ]{z.xiong@gsi.de}
\affiliation{GSI Helmholtzzentrum {f\"ur} Schwerionenforschung, Planckstra{\ss}e 1, D-64291 Darmstadt, Germany}

\author{Gabriel Mart\'{i}nez-Pinedo}
\affiliation{GSI Helmholtzzentrum {f\"ur} Schwerionenforschung, Planckstra{\ss}e 1, D-64291 Darmstadt, Germany}
\affiliation{Institut {f\"ur} Kernphysik (Theoriezentrum), Fachbereich Physik, Technische Universit{\"a}t Darmstadt, Schlossgartenstra{\ss}e 2, D-64289 Darmstadt, Germany}
\affiliation{Helmholtz Forschungsakademie Hessen {f\"ur} FAIR, GSI Helmholtzzentrum {f\"ur} Schwerionenforschung, Planckstra{\ss}e 1, D-64291 Darmstadt, Germany}

\author{Oliver Just}
\affiliation{GSI Helmholtzzentrum {f\"ur} Schwerionenforschung, Planckstra{\ss}e 1, D-64291 Darmstadt, Germany}
\affiliation{Astrophysical Big Bang Laboratory, RIKEN Cluster for Pioneering Research, 2-1 Hirosawa, Wako, Saitama 351-0198, Japan}

\author{Andre Sieverding}
\affiliation{Max Planck Institute for Astrophysics, Karl-Schwarzschild-Stra{\ss}e 1, D-85748 Garching, Germany}
\date{\today}
\maketitle
\graphicspath{{./}{./figures/}}

\setcounter{equation}{0}
\setcounter{figure}{0}
\setcounter{table}{0}
\renewcommand{\thetable}{S\arabic{table}}
\renewcommand{\theequation}{S\arabic{equation}}
\renewcommand{\thefigure}{S\arabic{figure}}
\renewcommand{\thesection}{S\arabic{section}}
\onecolumngrid

In this Supplemental Material, we provide additional information about the reaction dynamics of the $\nu r$-process.
We also compare the outflow conditions predicted by a neutron star merger simulation to the ones explored in our parametric survey.

\section{Reaction dynamics}

\begin{figure}[hbt]
    \centering
    \includegraphics[height=2.2in]{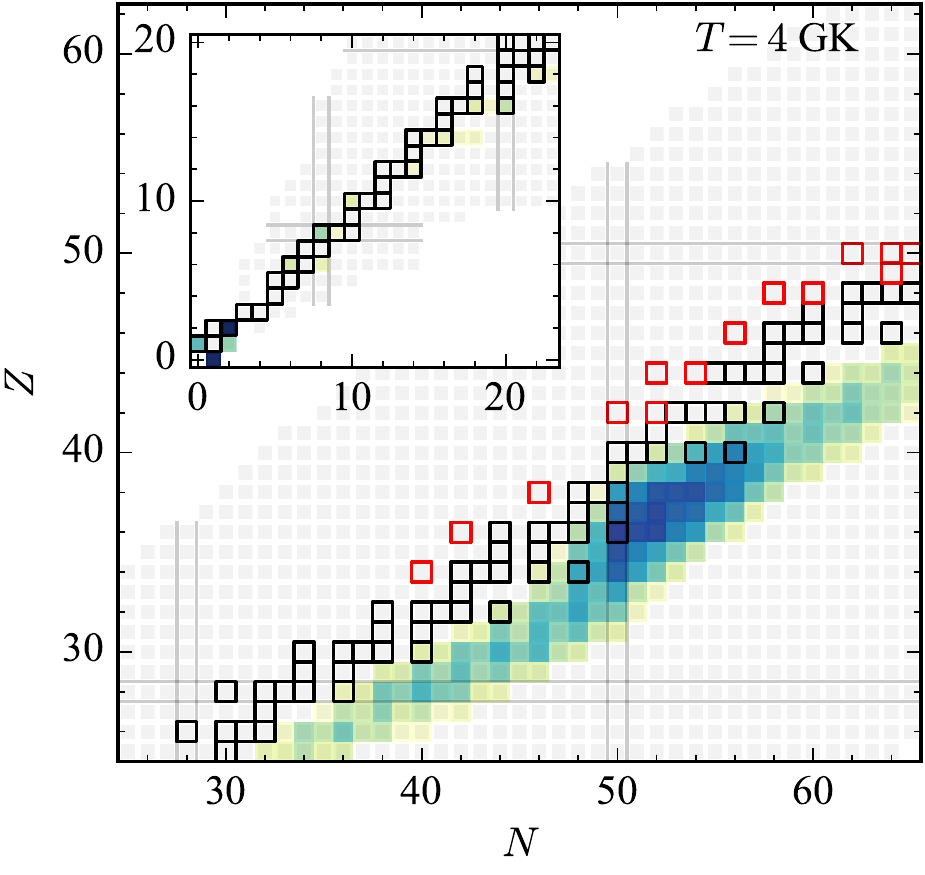}
    \includegraphics[height=2.2in]{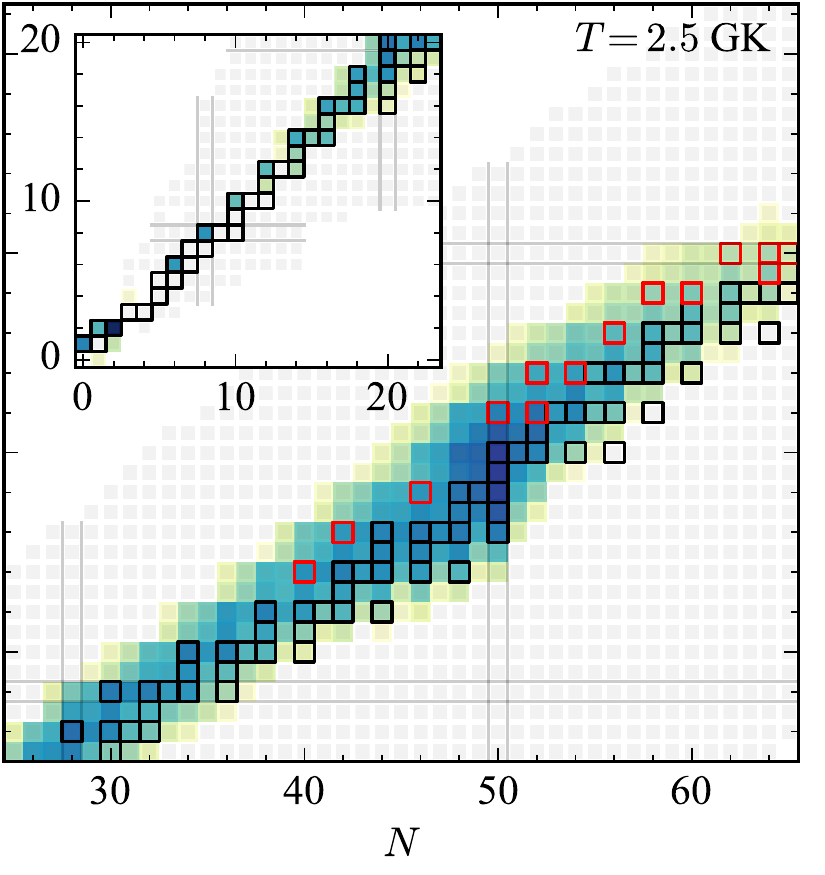}
    \includegraphics[height=2.24in]{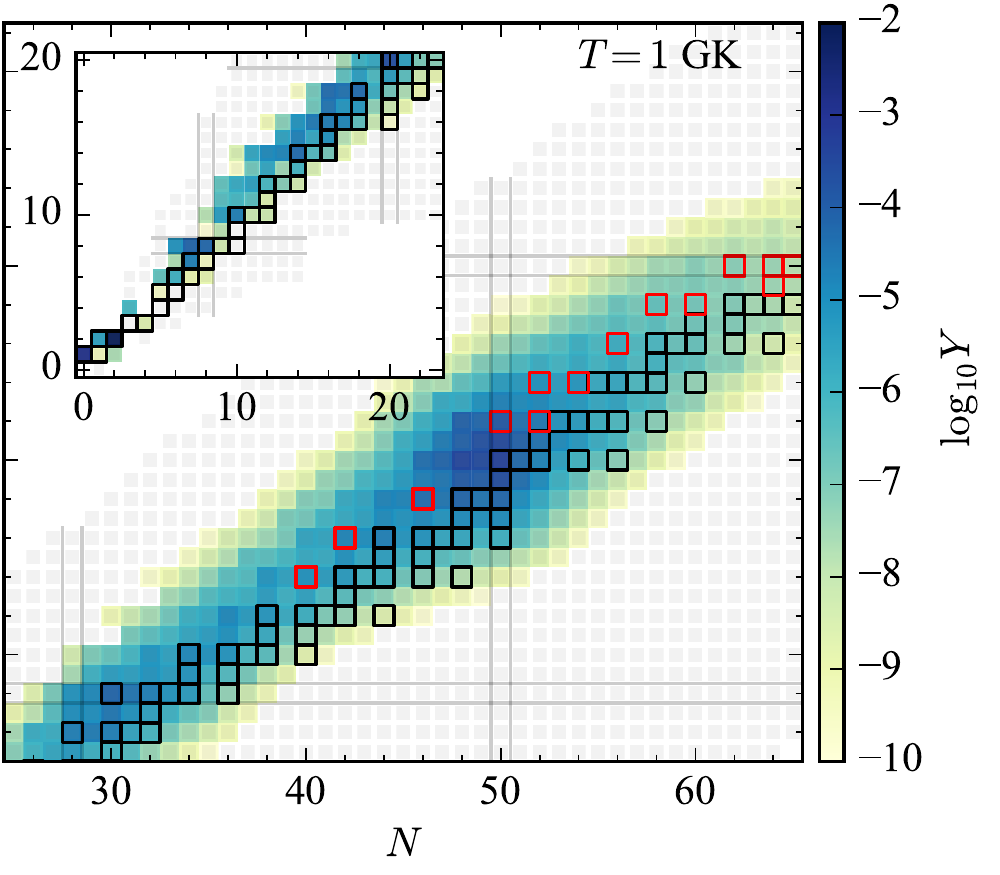}
    \caption{\label{fig:nz_plane}Abundance distributions in $N$-$Z$ plane at different stages for case I: synthesis of neutron-rich seeds (4~GK), neutron capture freeze-out ($\sim 2.5$~GK), and establishment of new equilibrium (1~GK). The light-element sectors are shown in smaller panels. Red- and black-edged squares are for $p$-nuclei and other stable nuclei, respectively. Double gray lines indicate the magic numbers.}
\end{figure}

Because the basic dynamics to produce $p$-nuclei is similar for all cases, we focus on the parameterized trajectory case I as it significantly coproduces $^{92,94}$Mo, $^{96,98}$Ru, and $^{92}$Nb.
Figure~\ref{fig:nz_plane} shows the abundance distributions in the $N$-$Z$ plane at different stages.
At 4~GK, the build-up of seeds is finished and neutron captures start to operate. The abundance distribution is dominated by neutron-rich nuclei with $N\sim 50$. 
Neutrino absorption operates dominantly on the most neutron-rich nuclei with larger charged-current cross sections and speeds up the build up of heavier elements and the depletion of neutrons.
At $2.5$~GK after the exhaustion of neutrons, charged-current (CC) neutrino absorption reactions move the abundance distribution across the stability line, which enables the production of $p$-nuclei and other neutron-deficient nuclei.
The neutrino reactions produce neutrons, which are recaptured by nuclei by $(n,\gamma)$ and $(n,p)$ reaction close to the magic number $N=50$.
$\nu$-A reactions and neutron captures reach an equilibrium with a broad abundance distribution at around $1$~GK that extends from neutron-deficient to neutron-rich nuclei. 
Unstable neutron-deficient nuclei eventually undergo $\beta^+$ decays and electron captures and are converted to $p$-nuclei and $s$-process nuclei.

\begin{figure}[hbt]
    \centering
    \includegraphics[width=0.32\columnwidth]{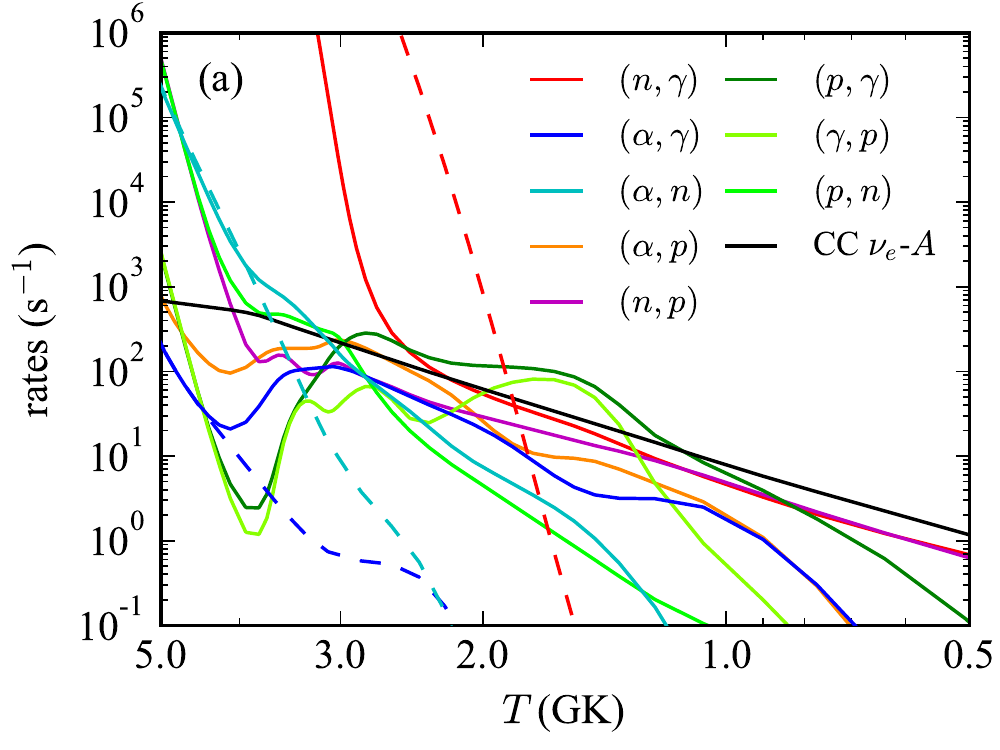}
    \includegraphics[width=0.32\columnwidth]{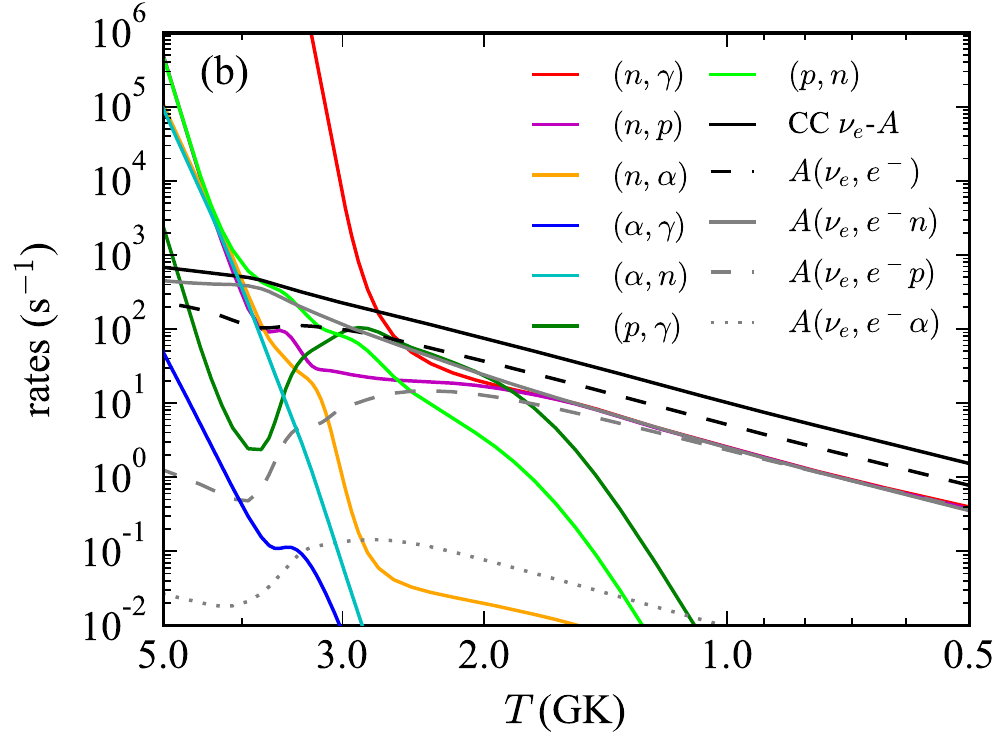}
    \includegraphics[width=0.32\columnwidth]{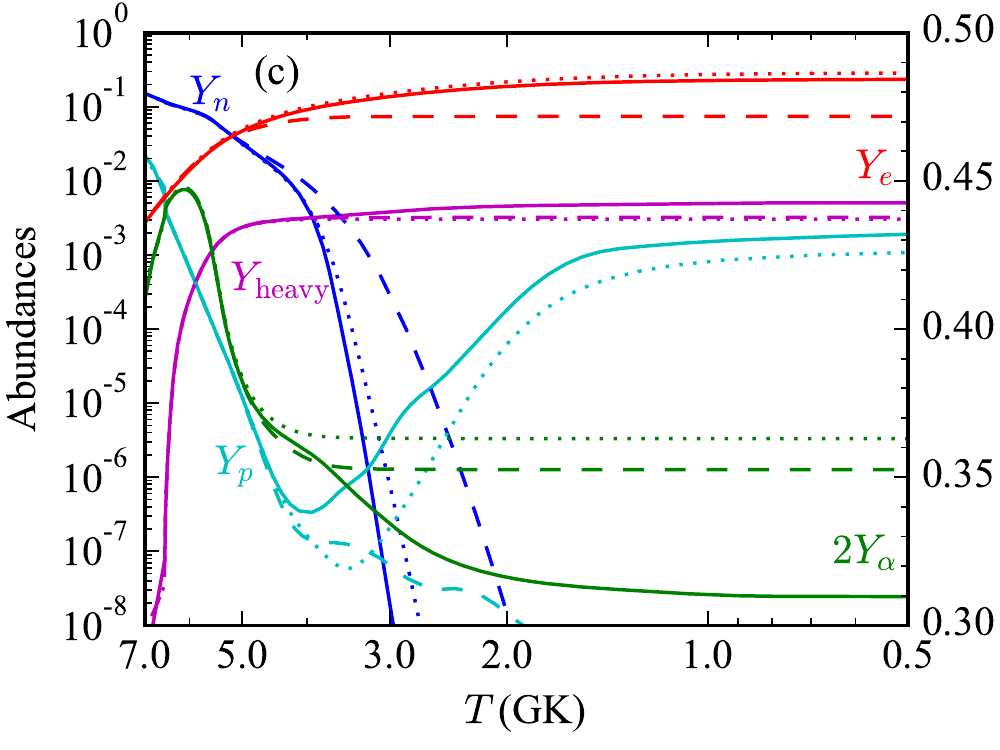}
    \caption{\label{fig:rates_nuclear_dynamics}Evolution of interaction rates averaged over all heavy nuclei ($A>4$) (a) and just nuclei with $A>80$ as in Eq.~\ref{eq:rate_Agt80} (b). Dashed and solid curves in (a) are for the cases without and with $\nu$-A reactions, respectively. (c) compares the abundances in the calculations without $\nu$-A reactions (dashed), with all $\nu$-A reactions (solid), and including $\nu$-A reactions except neutrino captures on $\alpha$-particles (dotted). All plots are for case I.}
\end{figure}

While the dynamics described above affects mostly the heaviest isotopes in the distribution, $\nu$-A interactions allow for additional reaction flows that affect the abundance pattern at lower mass numbers.
While these reactions do not lead to any substantial yield from the point of view of galactic chemical evolution, we describe their impact in the following as they represent a variety of nucleosynthesis scenarios that may not be realized in other environments. 
Initially, at $4$~GK, apart from $\alpha$ particles the abundance of nuclei is concentrated in the region $A\sim 70$--80 with a tail to lower masses that reaches up to $A=40$.
Between $A=40$ and $\alpha$-particles the abundance of nuclei is negligible. 
Neutrino spallation reactions on $^{4}$He continuously produce $^{3}$H and $^3$He that are quickly consumed by further $\alpha$-capture reactions.
This results in a continuous flow of material from $^4$He to heavier nuclei.
Interestingly, this allows to bridge the instability gaps at $A=5$ and $A=8$ by two-body reactions instead of the triple-alpha reaction.
How heavy nuclei are subsequently produced depends on temperature and the presence of neutrons.
At temperatures above $2.5$~GK, while neutron-captures dominate, the $\alpha$-process~\cite{Witti.Janka.Takahashi:1994,woosley1992alpha} brings material to the region between $A=40$ and 80 that is further processed by neutron capture reactions (see left panel of Fig.~\ref{fig:nz_plane}).
After the freeze-out of neutron captures, CC neutrino-nucleus reactions drive the distribution of heavy nuclei to and beyond the beta-stability line.
The reaction flow connecting $^{4}$He to heavier nuclei now proceeds by a sequence of reactions  similar to the $\alpha p$-process operating in accreting neutron stars~\cite{schatz2006x} up to $A=40$.
It is followed by a sequence of $(p,\gamma)$ and $(n,p)$ reactions that resembles the $\nu p$-process~\cite{frohlich2006neutrino} and reaches up to $A\sim 80$ (see the middle panel of Fig.~\ref{fig:nz_plane} that shows the distribution for material at 2.5~GK for case I).
This is manifested by the dominance of the $(p,\gamma)$ rate shown in Fig.~\ref{fig:rates_nuclear_dynamics} that is almost in equilibrium with the $(\gamma,p)$ rate in the temperature range 2--1~GK, and the fact that the $(\alpha,p)$ rate is larger than all other $\alpha$-capture reactions.
As the temperature decreases below 1~GK, the coulomb barriers above $Z=20$ become too large and the reaction flow stops at nuclei around $A=40$.
This produces an increase of the abundance of nuclei between $A=20$ and $A=40$ that can be seen by comparing the insets in the middle and right panels of Fig.~\ref{fig:nz_plane}.
$^{14}$O is produced during this phase by the hot CNO cycle.
The final abundances of $^{12}$C and $^{16}$O change from $1.7 \times 10^{-6}$and $2.6\times10^{-9}$ without $\nu$-A reactions to $4.9\times10^{-5}$ and $8.5\times10^{-5}$ with $\nu$-A reactions, respectively.
In addition, rare isotopes like $^7$Li and $^{11}$B are only produced when $\nu$-A reactions are considered and reach abundances of $1.4 \times 10^{-6}$ and $7.9 \times 10^{-6}$.

To separate the reaction dynamics operating  in the light and heavy elements, we show in Fig.~\ref{fig:rates_nuclear_dynamics}(b) rates averaged over nuclei with $A>80$ using the definition:
\begin{equation}
    \lambda_{I}^{A>80} = \frac{\sum_{A>80} \lambda_{I}(A,Z) Y({A,Z})}{\sum_{A>80} Y({A,Z})}
    \label{eq:rate_Agt80}
\end{equation}
for each particular reaction $I$.
Compared to the $(n,\gamma)$ and $\nu$-A reactions, the rates of $\alpha$-particle captures and $(n,\alpha)$ reactions are almost negligible for heavy nuclei with $A>80$.
The $(p,\gamma)$ and $(p,n)$ reactions have comparable rates (within one order of magnitude) to that of $(n,\gamma)$ at $T\sim 2$--3~GK, while they rapidly decay at lower temperature.
With more neutron-deficient nuclei produced at $T<2$~GK, the $(n,p)$ rate becomes almost identical to that of $(n,\gamma)$, which is dominated by nuclei with $A\sim 80$.

Figure~\ref{fig:rates_nuclear_dynamics}(b) also compares different spallation channels in CC $\nu_e$-A reactions.
At high temperatures $>3$~GK, $A(\nu_e,e^- n)$ and $A(\nu_e,e^-)$ channels have similar rates within one order of magnitude.  During the phase where neutron-captures dominate only the total CC $\nu_e$-A cross section is relevant.  
After the freeze-out of neutron captures, the CC $\nu_e$-A reaction, which leaves the mass number unchanged, is dominant for nuclei with $A>80$. As  the material moves to and beyond the beta-stability line, the rate of $A(\nu_e,e^- p)$ increases and becomes similar to the one of $A(\nu_e,e^- n)$.
The $A(\nu_e,e^- \alpha)$ channel has the least importance throughout the $\nu r$-process.
In cases II and III (not shown in the plot), the rates $\lambda^{A>80}$ for $(n,p)$, $(p,\gamma)$, $(p,n)$, and $A(\nu_e,e^- p)$ become subdominant, reduced by more than one order of magnitude compared to the rates of $(n,\gamma)$, $A(\nu_e,e^-)$, and $A(\nu_e,e^- n)$ reactions as the abundance distribution extends to higher mass numbers.

To separate the impact of the light-sector activity from the $\nu r$-process, we also perform calculations that include all $\nu$-A reactions except for those on $\alpha$-particles.
The final abundance of $p$-nuclei remains almost the same as when neutrino-$\alpha$ interactions are included. 
The evolution of $Y_e$ and $Y_n$, shown in dotted curves of Fig.~\ref{fig:rates_nuclear_dynamics}(c), is also similar to the case with all $\nu$-A reactions.
The major difference is that the light-sector activity is strongly suppressed so that $Y_{\rm heavy}$ and $Y_\alpha$ are almost constant below 4~GK.
Interestingly, the asymptotic value of $Y_\alpha$ is slightly higher than the one without $\nu$-A reactions, due to the contribution of $A(\nu_e,e^- \alpha)$ and $A(\nu,\nu\alpha)$  spallation reactions on heavy nuclei.
The proton abundance $Y_p$ behaves similarly to the case without $\nu$-A reactions before the freeze-out of neutron captures at $T\sim 4$~GK, because the supply of protons from neutrino spallation reactions on $\alpha$-particles is missing.
At lower temperatures $Y_p$ increases to a similar value as in the case with all $\nu$-A reactions because protons are continuously produced by a combination of neutrino spallation reactions [$A(\nu_e,e^- p)$, $A(\nu,\nu p)$], $(n,p)$ reactions, and $(\alpha,p)$ reactions.

\section{Outflow conditions in a merger simulation}
\begin{figure}[hbt]
	\centering
    \includegraphics[width=0.5\columnwidth]{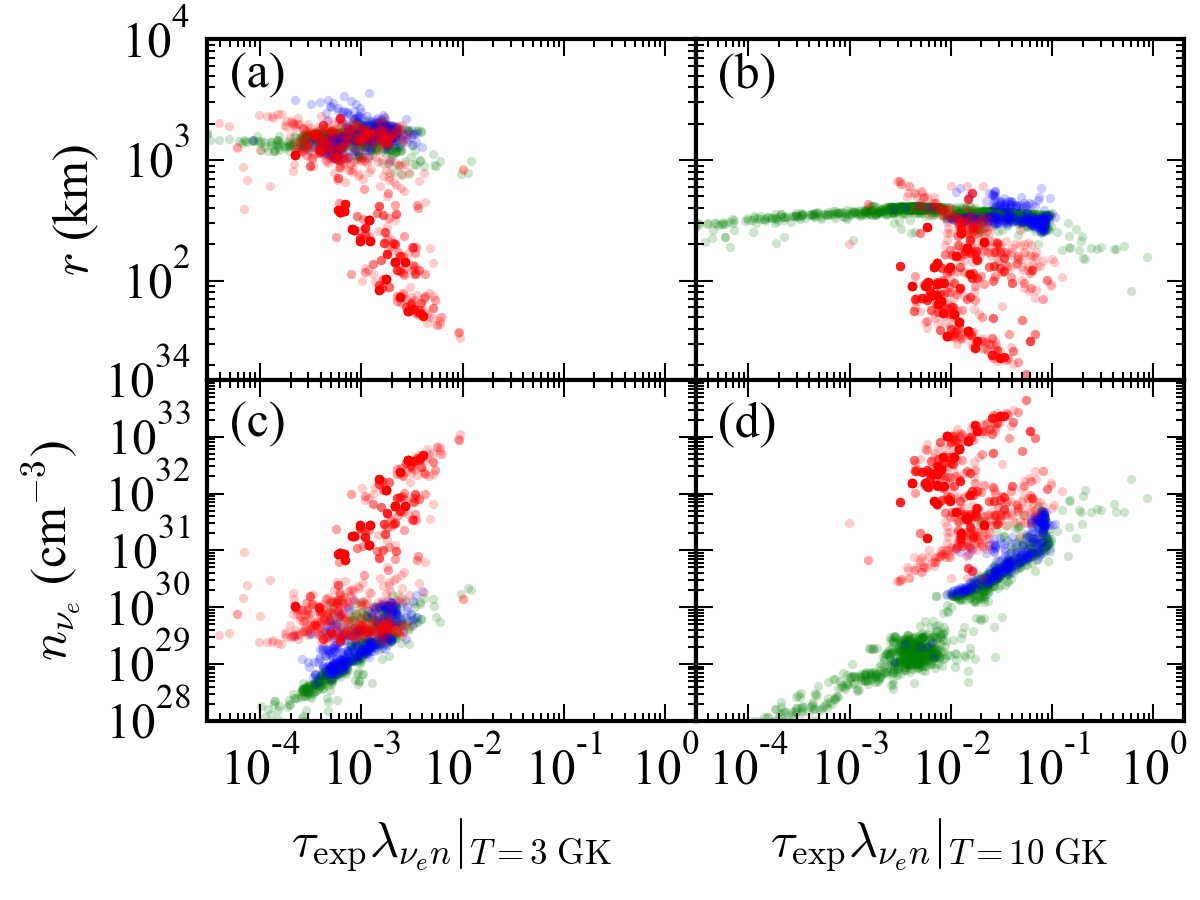}
    \caption{\label{fig:simulation}Neutrino exposure, radius, and neutrino density of outflow tracers at $T=3$~GK (left panels) and 10~GK (right panels) for the dynamical (red), neutron-star-torus (blue), and black-hole-torus (green) ejecta in the simulation model sym-n1-a6, respectively.}
\end{figure}

We consider the outflow conditions obtained for tracer particles in a hydrodynamic simulation of a binary neutron star merger, namely model sym-n1-a6 of Ref.~\cite{just2023end}. In Fig.~\ref{fig:simulation}, the neutrino exposure, radius, and neutrino density at temperatures of $T=3$~GK and $T=10$~GK are shown using colors to distinguish the three main ejecta components: dynamical ejecta (red), neutron-star torus ejecta (blue), and black-hole torus ejecta (green).

At $T=3$~GK, none of the tracers in this particular model experience neutrino exposures high enough (i.e. $\tau_{\text{exp}}\lambda_{\nu_e n}>0.1$) to enable the $\nu r$-process. Somewhat counterintuitively, the neutrino-driven wind blown off the neutron-star surface, which is sampled by a subset of the blue tracers, does not exhibit significantly higher neutrino exposure for a given temperature than the other ejecta components. This is because thermally driven neutrino winds tend to be overall hotter than the outflows driven by other mechanisms, meaning that a given temperature is reached only at a relatively large radius, where the neutrino flux density is relatively low. For this reason one may suspect that magnetically-driven outflows (which are not described by the considered simulation model) could be more promising to enable the $\nu r$ process.

Considering earlier expansion times with a slightly higher temperature of $T=10$~GK (right columns in Fig.~\ref{fig:simulation}), one can see how close the conditions already are to activating the $\nu r$ process: The neutrino exposures reach values of 0.1 and higher for tracers of all three ejecta components, with radii ranging from $\sim 100$~km to $\sim 600$~km. This suggests a strong sensitivity to the thermal conditions and calls for hydrodynamical models that predict temperatures with high accuracy. Hence, given the significant uncertainties of currently available simulation models of neutron-star mergers and core-collapse supernovae -- e.g. because of missing neutrino- or magnetic field related physics processes and limited numerical resolution -- it will be challenging to conclusively identify or exclude the operation of the $\nu r$ process in hydrodynamical models, particularly because this process may operate only in a very small fraction of the ejecta.

\bibliography{supplemental.bbl}